\documentclass[aps,prl,reprint,twocolumn,superscriptaddress]{revtex4-1}
\usepackage{amsmath}
\usepackage{amsfonts}
\usepackage{amssymb}
\usepackage{amsthm}
\usepackage{graphicx}
\usepackage{CJK}
\usepackage{times}
\usepackage{mathptmx}
\usepackage[colorlinks, linkcolor=blue, urlcolor=blue, anchorcolor=blue, citecolor=blue]{hyperref}
\usepackage{mathtools}
\usepackage{lipsum}
\usepackage{array}
\usepackage[table]{xcolor}
\definecolor{mygray}{gray}{.9}

\begin{document}


\title{Chiral domain wall dynamics in magnetic heterostructures with bulk Dzyaloshinskii-Moriya interactions}



\author{Mei Li}
\affiliation{College of Physics Science and Technology, Yangzhou University, Yangzhou 225002, People's Republic of China}
\author{Bin Xi}
\affiliation{College of Physics Science and Technology, Yangzhou University, Yangzhou 225002, People's Republic of China}
\author{Jie Lu}
\email{lujie@yzu.edu.cn}
\affiliation{College of Physics Science and Technology, Yangzhou University, Yangzhou 225002, People's Republic of China}
\author{Yongjun Liu}
\email{yjliu@yzu.edu.cn}
\affiliation{College of Physics Science and Technology, Yangzhou University, Yangzhou 225002, People's Republic of China}


\date{\today}

\begin{abstract}
In this work, dynamics of chiral domain walls in long and narrow 
magnetic heterostructures based on non-centrosymmetric chiral magnets with 
bulk Dzyaloshinskii-Moriya interactions (DMI) and perpendicular magnetic anisotropy  
is systematically investigated.
The driving forces can be out-of-plane magnetic fields and in-plane currents,
correspondingly both steady and precessional flows are considered.
Their dividing points (the Walker critical field and current density) are obtained as functions 
of bulk DMI strength ($D_{\mathrm{b}}$) and the ratio ($\kappa$) of total (crystalline plus shape) 
anisotropy in the hard axis over that in the easy one.
When far beyond Walker breakdown, the dependence curve of wall velocity on external in-plane bias field 
takes parabolic shape around the compensation point where the total in-plane field disappears.
The center shift is determined by $D_{\mathrm{b}}$, $\kappa$, and the wall's topological charge,
thus can be used to measure the bulk DMI strength in chiral magnets.

\end{abstract}


\maketitle


\section{I. Introduction} 
Magnetic solitons with spatial localization and topological protection
have attracted intense attentions in the past decades due to
both academic and industrial interests.
Recently, discussions about chiral magnetic solitons stabilized by the 
Dzyaloshinskii-Moriya interaction (DMI) become extraordinarily active.
The most common examples are chiral domain walls (DWs)\cite{Blugel_PRB_2008,EPL_100_57002,ChenG_PRL_2013,ChenG_NatCommun_2017,Linder_PRB_2017,jlu_NJP_2019}, 
skyrmions/antiskyrmions\cite{Boni_Science_2009,Nagaosa_Nature_2010,Hoffmann_PhysRep_2017,XiB_NanoLett_2019,ZhangXC_JPCM_2020} 
and bimerons\cite{Ezawa_PRB_2011,Batista_PRB_2015,Tretiakov_PRB_2019,ShenLC_PRL_2020,ZhangXC_PRB_2020}, etc.
Historically, the bulk DMI (bDMI) was first proposed which should phenomenologically
include an odd term of the spatial gradient
of magnetization as a result of being an antisymmetric exchange coupling\cite{Dzyaloshinsky}.
From the microscopic viewpoint, bDMI comes from the generalization of Anderson's superexchange
theory in the presence of spin-orbit coupling\cite{Moriya}.
Experimentally, it was first proposed to reside in 
chiral magnets with non-centrosymmetric B20 structure\cite{Boni_Science_2009,Nagaosa_Nature_2010,Wiesendanger_Nature_2007,Chien_PRL_2012}.
In addtion, the non-collinear magnetic structures observed recently in several Heusler
compounds also suggest its possible existence therein\cite{Meshcheriakova_PRL_2014,Chadov_NatCommun_2016}.
Magnetic heterostructures based on these novel materials with bDMI should open new possibility
of future spintronics devices with chiral magnetic solitons serving as 
carriers of information recording and transmission.

In many proposed experiments, magnetic heterostructures are prepared on heavy-metal substrates\cite{Pizzini_JPCM_2015,Pizzini_PRB_2016,Pizzini_EPL_2016,Pizzini_APL_2017,Pizzini_PRB_2019,Jung_srep_2016,Perna_NanoLetters_2018,Pizzini_PRL_2018,Fukami_APL_2019,Choe_PRB_2019,Ohno_nphys_2016,Parkin_NC_2018,Klaui_PRL_2018,Hrabec_Nanotechnology_2019,Lau_AIPAdvances_2019}. 
In these setups, the interfacial DMI (iDMI), spin Hall and Rashba spin-orbit torques emerge 
thus complicatedly manipulate the motion of chiral DWs in the primary magnetic layer under external currents\cite{Ohno_nphys_2016,Parkin_NC_2018,Klaui_PRL_2018,Hrabec_Nanotechnology_2019,Lau_AIPAdvances_2019,PBH_PRB_2020}.
To explore the effects of pure bDMI on chiral-DW dynamics,
in this work we focus on long and narrow magnetic heterostructures in which chiral magnets 
with perpendicular magnetic anisotropy (PMA)
are sandwiched between normal insulating substrates and caplayers.
Once nucleated, chiral DWs can be driven to move longitudinally
by either out-of-plane magnetic fields or in-plane currents.
In principle, both steady and precessional flows can emerge.
However their dividing point, the Walker limit, will be manipulated by the bDMI subtly.
When far beyond the Walker breakdown, the dependence of wall velocity on in-plane magnetic 
bias fields is explored. 
The resulting curves take parabolic shapes around the compensation point where the total (external plus internal)
in-plane field disappears.
Accordingly, the bDMI strength can be obtained directly from the center shifts of these curves.

The rest of this paper is organized as follows. 
In Sec. II the system set up and its modelization are briefly introduced. 
Also, the Lagrangian-based collective coordinate model adopted is presented.
Then the chirality of static walls is investigated in Sec. III.
After that, the field-driven and current-driven dynamics of chiral DWs
are systematically studied in Sec. IV and V, respectively.
Finally, concluding remarks are provided in the last section.

\section{II. Model and preparation} 
Generally, the magnetic free-energy density $\mathcal{E}_{0}$ of the chiral magnet in a heterostructure 
(see Fig. 1) includes four parts: 
the exchange part $\mathcal{E}_{\mathrm{ex}}=A(\mathbf{\nabla}\mathbf{m})^2$ ($A$ and $\mathbf{m}$ 
being the exchange stiffness and magnetization unit vector, respectively), 
the Zeeman part $\mathcal{E}_{\mathrm{Z}}=-\mu_0 M_s\mathbf{m}\cdot\mathbf{H}_a$ with
the total external field $\mathbf{H}_a=\mathbf{H}_z+\mathbf{H}_{\perp}$ and the saturation magnetization $M_s$,
the anisotropy part $\mathcal{E}_{\mathrm{ani}}=(\mu_0 M_s^2/2)(-k_{\mathrm{E}}m_z^2+k_{\mathrm{H}}m_y^2)$ where $k_{\mathrm{E}}$ ($k_{\mathrm{H}}$) is the total (crystalline plus shape) anisotropy coefficient in easy (hard) axis (in this work we consider PMA, which is the most common in chiral magnets), and the bDMI contribution  $\mathcal{E}_{\mathrm{b}}=D_{\mathrm{b}}\mathbf{m}(\mathbf{r})\cdot[\mathbf{\nabla}\times\mathbf{m}(\mathbf{r})]$
with $D_{\mathrm{b}}$ being the bDMI strength\cite{Bak_JPC_1980}.
The corresponding bDMI-induced effective field then reads
$\mathbf{H}_{\mathrm{b}}(\mathbf{r})=-2D_{\mathrm{b}}(\mathbf{\nabla}\times \mathbf{m})/(\mu_0 M_s)$.

\begin{figure} [htbp]
	\centering
	\includegraphics[width=0.44\textwidth]{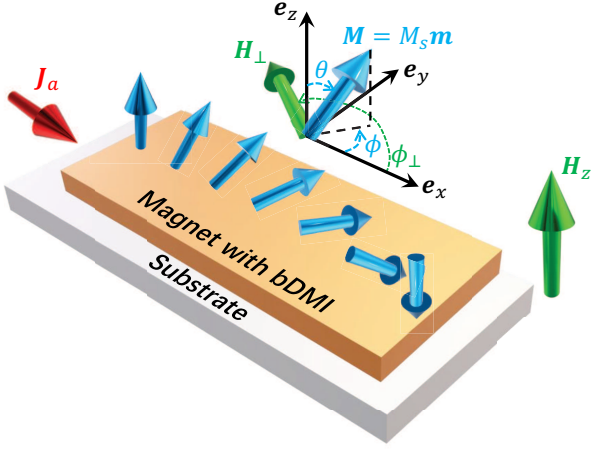}
	\caption{(Color online) Sketch of a narrow-strip-shaped heterostructure in which a magnet with bDMI and 
		PMA is prepared on a nonmagnetic substrate.
		The magnetization $\mathbf{M}=M_s \mathbf{m}$ can be fully described by its polar and
		azimuthal angles ($\theta$ and $\phi$) .
		An ``$\uparrow\downarrow$" wall is driven to move in $x-$direction 
		by either an external in-plane current density $\mathbf{J}_a$ or an out-of-plane field $\mathbf{H}_z$. 
		In the meantime, an in-plane bias field, $\mathbf{H}_{\perp}=H_{\perp}(\cos\phi_{\perp}\mathbf{e}_x+\sin\phi_{\perp}\mathbf{e}_y)$,
		is applied to manipulate the wall's behavior.
	}\label{fig1}
\end{figure}

Under out-of-plane magnetic fields and in-plane currents, the Lagrangian $\mathcal{L}$ of this chiral magnet is
\begin{equation}\label{Lagrangian}
\frac{\mathcal{L}}{\mu_0 M_s^2}=-\frac{\cos\theta}{\gamma M_s}\frac{\partial\phi}{\partial t}-\frac{B_J\phi}{\gamma M_s}\frac{\partial\cos\theta}{\partial(\hat{\mathbf{J}}\cdot\mathbf{r})}-\frac{\mathcal{E}_{0}}{\mu_0 M_s^2},
\end{equation}
with the dissipative functional
\begin{equation}\label{Damping_functional}
\frac{\mathcal{F}}{\mu_0 M_s^2}=\frac{\alpha}{2\gamma M_s}\left\{\left[\frac{\partial}{\partial t}-\frac{\beta B_J}{\alpha}\frac{\partial}{\partial(\hat{\mathbf{J}}\cdot\mathbf{r})}\right]\mathbf{m}\right\}^2
\end{equation}
describing the various damping processes\cite{Boulle_PRL_2013,He_EPJB_2013,jlu_PRB_2019,jlu_PRB_2020,jlu_JMMM_2021,jlu_PRB_2021}.
Here $\theta(\mathbf{r},t)$ and $\phi(\mathbf{r},t)$ are the polar and 
azimuthal angles of $\mathbf{m}(\mathbf{r},t)$, respectively.
$\alpha$ is the damping constant and $\beta$ is the nonadiabatic spin-transfer torque (STT) coefficient.
$\gamma=\mu_0\gamma_e$ with $\mu_0$ and $\gamma_e$ being the vacuum permeability and electron
gyromagnetic ratio, respectively.
$B_J=\mu_{\mathrm{B}}Pj_a/(e M_s)$, in which $\mu_{\mathrm{B}}$ is the Bohr magneton 
and $e(>0)$ is the absolute electron charge. 
$j_a$ (with unit vector $\hat{\mathbf{J}}$) is the current density flowing longitudinally 
through the chiral magnet with polarization $P$.

The magnetzation dynamics of the chiral magnet is described by the generalized Lagrangian equation,
\begin{equation}\label{EL_equation}
\frac{d}{d t}\left(\frac{\delta\mathcal{L}}{\delta \dot{X}}\right)-\frac{\delta \mathcal{L}}{\delta X}+\frac{\delta\mathcal{F}}{\delta \dot{X}}=0,
\end{equation}
where an overdot means $\partial/\partial t$ and $X$ is any related coordinate.
To explore collective behaviors, we use the Lagrangian-based collective coordinate model
which needs preset ansatz of DWs.
For long and narrow heterostructures, we take the quasi one dimensional (1D) Walker ansatz\cite{Slonczewski_1972,Thiaville_EPL_2005}
\begin{equation}\label{q_phi_Delta_ansatz}
\ln\tan\frac{\vartheta}{2}=\eta\frac{x-q(t)}{\Delta},\quad \phi=\varphi(t)
\end{equation}
in which $q$, $\Delta$, and $\varphi$ are wall center position, wall width and in-plane magnetization angle, respectively.
$\eta=+1(-1)$ is the topological wall charge which corresponds to ``$\uparrow\downarrow(\downarrow\uparrow)$" wall.
For narrow-strip geometry as shown in Fig. 1, the $\mathbf{e}_x$ and $\mathbf{e}_y$ axes respectively 
indicate the ``longitudinal (L)" and ``transverse (T)" directions.
For this quasi one-dimensional system, the in-plane component of the effective field from bDMI 
becomes $\mathbf{H}_{\mathrm{b}}(x)=2D_{\mathrm{b}}(\nabla_x m_z)\mathbf{e}_y/(\mu_0 M_s)$.
Clearly, $\mathbf{H}_{\mathrm{b}}$ has transverse component proportional to 
$\nabla_x m_z$, which is reversed under wall charge reversal $\eta\rightarrow -\eta$.
By setting $X=q$, $\varphi$, $\Delta$ and integrating the resulting
equations along longitudinal direction ($\int_{-\infty}^{+\infty}dx$), 
the following closed dynamical equation set is obtained,
\begin{equation}\label{Full_dynamical_equation_bDMI}
\begin{split}
(1+\alpha^2)\dot{\varphi}=&\gamma H_z+(\alpha-\beta)\frac{\eta B_J}{\Delta}-\frac{\alpha\pi\gamma}{2}\left[\frac{k_{\mathrm{H}}M_s}{\pi}\sin 2\varphi\right.  \\
&\left.  +H_{\perp}\sin(\varphi-\phi_{\perp})+\frac{\eta D_{\mathrm{b}}\cos\varphi}{\mu_0 M_s\Delta}\right], \\
\dot{q}=& -\frac{\eta\Delta}{\alpha}\dot{\varphi}+\frac{\eta\Delta\gamma}{\alpha}H_z-\frac{\beta}{\alpha}B_J, \\
\frac{\alpha\pi}{6\gamma_0}\frac{\dot{\Delta}}{\Delta}=&\frac{2A}{\pi\mu_0 M_s \Delta^2}-\frac{M_s}{\pi}\left(k_{\mathrm{E}}+k_{\mathrm{H}}\sin^2\varphi\right)   \\
&  +H_{\perp}\cos(\varphi-\phi_{\perp}).
\end{split}
\end{equation}
These are all we need to proceed our investigation.

\section{III. Chirality of static walls} 
As the first step, the chirality of static walls selected by bDMI is explored.
When $H_z=0$ and $j_a=0$, the wall keeps static. 
Under the Walker profile and in the absence of in-plane bias fields, the total magnetic energy 
$E_0/S =\int_{-\infty}^{+\infty}\mathcal{E}_0[\mathbf{M}] dx=2A/\Delta +\mu_0 M_s^2\Delta (k_{\mathrm{E}}+k_{\mathrm{H}}\sin^2\varphi)+\eta\pi D_{\mathrm{b}}\sin\varphi$,
in which $S$ is the cross-sectional area of the chiral magnet.
For static walls, the last equation in Eq. (\ref{Full_dynamical_equation_bDMI}) provides
$\Delta=\Delta_0(1+\kappa\sin^2\varphi)^{-1/2}$ 
with $\Delta_0=\sqrt{2A/(\mu_0 k_{\mathrm{E}} M_s^2)}$ and $\kappa=k_{\mathrm{H}}/k_{\mathrm{E}}$.
Therefore, one has $E_0/S=2[2A\mu_0 M_s^2(k_\mathrm{E}+k_{\mathrm{H}}\sin^2\varphi)]^{1/2}+\eta\pi D_{\mathrm{b}}\sin\varphi$.
Obviously when bDMI is absent, either $\varphi=0$ or $\varphi=\pi$ provides the minimum of $E_0$
thus the wall takes N\'{e}el-type profile and no chirality is preferred.
However, as $D_{\mathrm{b}}$ appears, the minimization operation of $E_0/S$ provides,
\begin{widetext} 
\begin{equation}\label{Static_wall_chirality}
\frac{(E_0)_{\mathrm{min}}}{\pi S D_0}=
\begin{cases}
\sqrt{\left(1+\frac{1}{\kappa}\right)\frac{1}{\kappa}-\left(\frac{D_{\mathrm{b}}}{D_0}\right)^2\frac{1}{\kappa}}\quad  & \mathrm{at}\quad \sin\varphi_0=-\eta\mathrm{sgn}(D_{\mathrm{b}})\sqrt{\frac{(D_{\mathrm{b}}/D_0)^2}{(1+\kappa)-\kappa(D_{\mathrm{b}}/D_0)^2}} \quad \mathrm{for} \quad |D_{\mathrm{b}}|\le D_0, \\
\left(1+\frac{1}{\kappa}\right)-\frac{|D_{\mathrm{b}}|}{D_0}\quad & \mathrm{at}\quad \sin\varphi_0=-\eta\mathrm{sgn}(D_{\mathrm{b}}) \quad \mathrm{for} \quad |D_{\mathrm{b}}|>D_0,
\end{cases}
\end{equation}
\end{widetext} 
with $D_0\equiv (2 M_s k_{\mathrm{H}}/\pi)\sqrt{2A\mu_0/(k_{\mathrm{E}}+k_{\mathrm{H}})}$
and ``$\mathrm{sgn}$" being the sign function.
This means that for finite bDMI ($|D_{\mathrm{b}}|\le D_0$), the wall profile
is a mixture of N\'{e}el and Bloch types thus shows certain chirality preference through
non-zero $\left\langle m_y \right\rangle$.
While for sufficiently large bDMI ($|D_{\mathrm{b}}|> D_0$), chiral Bloch walls emerge.
This process is shown in Fig. 2 where $\eta=+1$ and $D_{\mathrm{b}}<0$ are taken as an example.

\begin{figure} [htbp]
	\centering
	\includegraphics[width=0.5\textwidth]{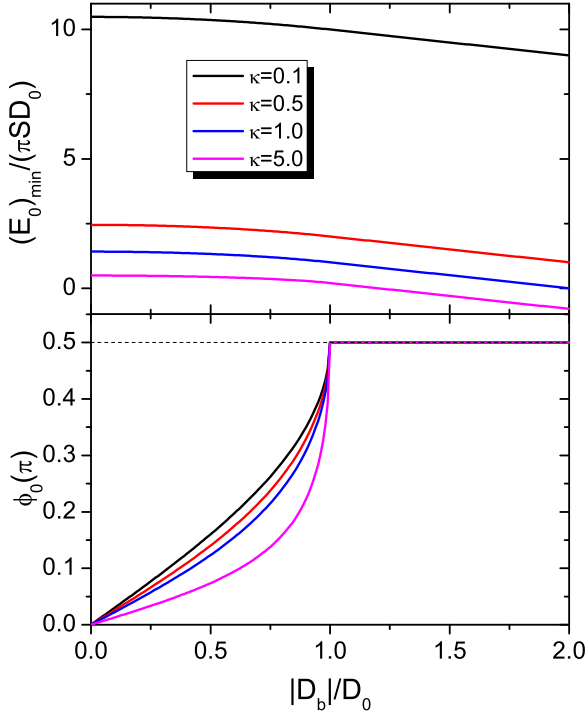}
	\caption{(Color online) Evolution of (a) minimum total magnetic energy and (b) the corresponding
		in-plane angle of chiral DWs as bDMI increases. Four typical ``hard-easy ratio" $\kappa$
		are presented. Obviously for a given $\kappa$, there exists a critical bDMI stength $D_0$.
		When $|D_{\mathrm{b}}|<D_0$ ($\ge D_0$), the wall shows partial (full) chirality.
	}\label{fig2}
\end{figure}

When finite out-of-plane fields and/or in-plane currents are applied, Eq. (\ref{Full_dynamical_equation_bDMI})
implies that there should be two dynamical modes: the steady-flow mode for small driving factors
and the precessional-flow mode for sufficiently large external stimuli.
The dividing point is the ``Walker field" or ``Walker current density", which will be manipulated by the bDMI.
In the following two sections, field- and current-driven dynamics of these chiral DWs
will be respectively investigated.

\section{IV. Field-driven dynamics} 
In this section, we focus on the chiral DWs dynamics under pure 
out-of-plane driving field $\mathbf{H}_z=H_z\mathbf{e}_z$. 
First in the absence of $\mathbf{H}_{\perp}$, the effects of bDMI on 
both Walker field and high-field wall behaviors are investigated.
Further manipulations of in-plane bias fields to
chiral walls' velocity under high $H_z$ are then analyzed which provides
applicable procedure of measuring the bDMI strength $D_{\mathrm{b}}$.

\subsection{IV.A Enlarged Walker field}
First we define several quantities for convenience.
They are: the anisotropy field in hard axis $H_{\mathrm{K}}=k_{\mathrm{H}}M_s$,
the original Walker field $H_{\mathrm{W}}^0=\alpha H_{\mathrm{K}}/2$,
the bDMI effective field strength $H_{\mathrm{b}}^0=D_{\mathrm{b}}/(\mu_0 M_s \Delta_0)$,
and the dimensionless coefficient $b=\eta\pi H_{\mathrm{b}}^0/H_{\mathrm{K}}$.
In the absence of $\mathbf{H}_{\perp}$ and $j_a$, the rigid-flow mode requires 
$\dot{\varphi}=0$ and $\dot{\Delta}=0$ which leads to
\begin{equation}\label{f_phi_definition}
\frac{H_z}{H_{\mathrm{W}}^0}=f(\varphi)\equiv b\cos\varphi\sqrt{1+\kappa\sin^2\varphi}+\sin 2\varphi.
\end{equation}
For fixed $b$ and $\kappa$, once the maximum absolute value of the function $f$, i.e. $|f(\varphi)|_{\mathrm{max}}$,
is found for $\varphi\in\left[0,2\pi \right)$, the new Walker field then reads
$H_{\mathrm{W}}=H_{\mathrm{W}}^0  |f(\varphi)|_{\mathrm{max}}$.

\hspace*{\fill} 

\begin{figure} [htbp]
	\centering
	\includegraphics[width=0.43\textwidth]{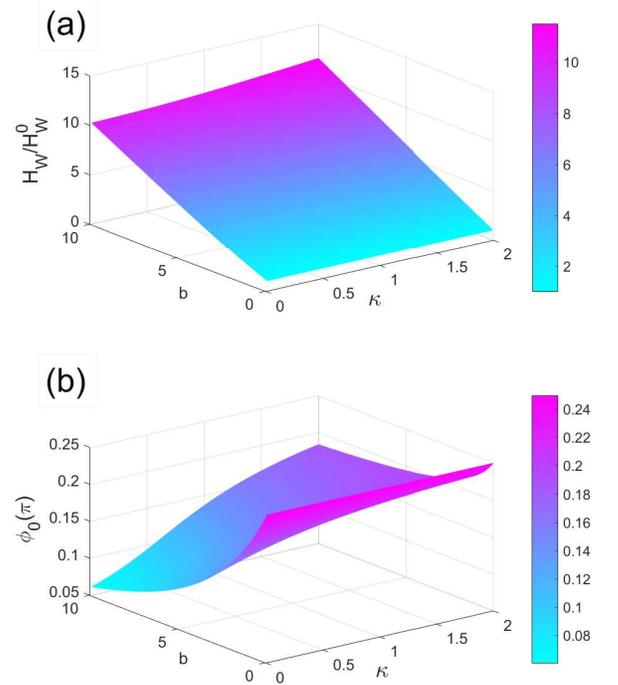}
	\caption{(Color online) (a) The new Walker field $H_{\mathrm{W}}$ in the presence of bDMI 
		and (b) the corresponding in-plane angle at which $H_{\mathrm{W}}$ is achieved as functions
		of the bDMI effective field strength ($b=\eta\pi H_{\mathrm{b}}^0/H_{\mathrm{K}}$) 
		and ``hard-easy ratio" $\kappa$. The data are calculated based on the results in Appendix A. 
	}\label{fig3}
\end{figure}

Before presenting the detailed results, several points need to be clarified:
(i) the sign of $b$ does not affect $|f(\varphi)|_{\mathrm{max}}$ since we always have 
$f(b,\varphi)=f(-b,\varphi+\pi)$. Therefore without loss of generality, we set $b>0$.
(ii) if $\kappa$ can be neglected, then after simple algebra we find that
at $\sin\varphi=(-b+\sqrt{b^2+32})/8$ the function $f$ reaches its maximum absolute value
$|f(\varphi,\kappa=0)|_{\mathrm{max}}=\sqrt{(-b^4+80b^2+128)+b(b^2+32)^{3/2}}/8\sqrt{2}$,
which recovers the result in one of our recent works\cite{jlu_NJP_2019}.
However, in real magnetic heterostructures, $\kappa$ is generally the order of 1.
After systematic calculus, the explicit form of $|f(\varphi)|_{\mathrm{max}}$ can be
obtained analytically (see Appendix A).
Based on it, the enlarged Walker field $H_{\mathrm{W}}$ (in the unit of $H_{\mathrm{W}}^0$) 
and the corresponding in-plane angle $\varphi_0$ where $H_{\mathrm{W}}$ is achieved
are calculated and plotted in Fig. 3(a) and 3(b), respectively.
As bDMI gradually increases, the Walker field is enlarged and its location, 
$\varphi_0$, decreases from $\pi/4$ to 0.
Obviously, the ``hard-easy ratio" $\kappa$ hardly affects $H_{\mathrm{W}}/H_{\mathrm{W}}^0$ 
however it strongly manipulates its location $\varphi_0$.
When $\kappa \gg 1$ the location of Walker field is nearly unchanged.
This can be easily understood from Eq. (\ref{f_phi_definition}) since now one has 
$f(\varphi)\sim (b\sqrt{\kappa}/2+1)\sin 2\varphi$, which achieves its maximum at $\varphi=\pi/4$.

For $|H_z|<H_{\mathrm{W}}$, one should first numerically solve Eq. (\ref{f_phi_definition}) 
to obtain the in-plane angle $\varphi(H_z)$ thus the wall width $\Delta=\Delta_0(1+\kappa\sin^2\varphi)^{-1/2}$.
Then the wall velocity reads $\dot{q}=\eta\Delta\gamma H_z/\alpha$.
In particular, given a certain chiral magnet with fixed shape (thus fixed $H_{\mathrm{b}}^0$ and $H_{\mathrm{K}}$):
(a) for a given wall topological charge $\eta$, when $\mathbf{H}_z\rightarrow -\mathbf{H}_z$, one has
$\varphi(H_z)\rightarrow \pi-\varphi(H_z)$ leading to unchanged wall width thus opposite wall velocity;
(b) for a given $\mathbf{H}_z$, the $\eta\rightarrow -\eta$ operation results in 
$\varphi(H_z)\rightarrow \pi+\varphi(H_z)$. Therefore the wall width is also unchanged and
eventually wall velocity is reversed.

\subsection{IV.B $\langle\dot{q}\rangle$ under $|H_z|\gg H_{\mathrm{W}}$}
When $|H_z|$ exceeds $H_{\mathrm{W}}$, the Walker breakdown takes place 
and the wall falls into precessional-flow mode.
Generally the breathing wall width ($\dot{\Delta}\ne 0$) has no explicit expression. 
In the simplest approximation, one can still take $\Delta=\Delta_0(1+\kappa\sin^2\varphi)^{-1/2}$.
By integrating the first equation in Eq. (\ref{Full_dynamical_equation_bDMI}) in a full circle, the
precessional period $T_0$ is obtained as
\begin{equation}\label{T_Hz_integration_form}
\bar{\gamma}T_0=\int_{0}^{2\pi}\frac{d\varphi/\sqrt{1+\kappa\sin^2\varphi}}{\frac{H_z-H_{\mathrm{W}}^0\sin 2\varphi}{\sqrt{1+\kappa\sin^2\varphi}}-\frac{\eta\alpha\pi}{2}H_{\mathrm{b}}^0\cos\varphi},
\end{equation}
where $\bar{\gamma}\equiv \gamma/(1+\alpha^2)$ and holds throughout this work.
When bDMI is absent, the integration can be easily calculated. 
As bDMI emerges, the situation becomes complicated. 
Generally, Eq. (\ref{T_Hz_integration_form}) has no explicit form.
However, in the high-field limit in which $|H_z|\gg \alpha |H_{\mathrm{b}}^0|(H_{\mathrm{W}}^0)$, by using the 
approximation $(1-\epsilon)^{-1}\approx 1+\epsilon+\epsilon^2$, Eq. (\ref{T_Hz_integration_form}) gives
\begin{equation}\label{T_Hz_with_2nd_order_terms}
T_0\approx\frac{2\pi}{\bar{\gamma}H_z}\left[1+\frac{\alpha^2\pi^2}{8}\left(1+\frac{\kappa}{4}\right)\left(\frac{H_{\mathrm{b}}^0}{H_z}\right)^2+\frac{\alpha^2}{8}\left(\frac{H_{\mathrm{K}}}{H_z}\right)^2\right].
\end{equation}
Therefore the time-averaged wall velocity is
\begin{equation}\label{v_average_Hz_with_2nd_order_terms}
\langle\dot{q}\rangle_0=\eta\alpha\Delta_0\bar{\gamma}H_z\frac{K_0}{2\pi}\left[1+\frac{\pi^2}{8}\left(1+\frac{\kappa}{4}\right)\left(\frac{H_{\mathrm{b}}^0}{H_z}\right)^2+\frac{1}{8}\left(\frac{H_{\mathrm{K}}}{H_z}\right)^2\right],
\end{equation}
in which $K_0$ is defined in Eq. (\ref{K_0_definition}).
If the velocity dependence on $H_z$ is the main concern, 
interestingly Eq. (\ref{v_average_Hz_with_2nd_order_terms}) can be reorganized as
$c(H_z-H_0)^2/H_z+d/H_z$, which is exactly the same as Eq. (9) of our early work in Ref. \cite{jlu_EPL_2009}.
Once again, the correctness of our original roadmap on field-driven DW dynamics is verified.
Alternatively, when we focus on the manipulation of bDMI on walls' drifting velocity, a parabolic $\langle\dot{q}\rangle_0\sim H_{\mathrm{b}}^0$ relationship
emerges which is similar to Eq. (9) in Ref. \cite{Choe_PRB_2019}. 
Note that our result here has two advantages: 
(a) here the ``$H_{\mathrm{W}}^0\sin 2\varphi$"-term has been preserved thus 
leading to the $(H_{\mathrm{K}})^2$ term which is missing in Ref. \cite{Choe_PRB_2019};
(b) the dependence of $\langle\dot{q}\rangle_0$ on the ``hard-easy ratio" $\kappa$ is fully revealed,
which has been totally neglected in most existing literatures.

\subsection{IV.C $\langle\dot{q}\rangle \sim H_{\perp}$ dependence under high $H_z$}
Next we turn on the in-plane bias field
$\mathbf{H}_{\perp}=H_{\perp}(\cos\phi_{\perp}\mathbf{e}_x+\sin\phi_{\perp}\mathbf{e}_y)$.
Generally the Walker field $H_{\mathrm{W}}$ will be further enlarged due to the ``pinning" effect of 
$\mathbf{H}_{\perp}$ to the in-plane angle $\varphi$, however the explicit form is mathematically
hopeless due to the mismatch between the symmetries of quadratic anisotropy and linear Zeeman energies.
In this subsection, we focus on $\langle\dot{q}\rangle \sim H_{\perp}$ dependence under
sufficiently large $H_z$ where chiral DWs take precessional motion.
Similarly, in the simplest approximation the wall width is expressed as $\Delta=\Delta_0(1+\kappa\sin^2\varphi)^{-1/2}$.
The period for a full circle is similar with that in Eq. (\ref{T_Hz_integration_form}) 
except for an additional $-\frac{\alpha\pi}{2}H_{\perp}\sin(\varphi-\phi_{\perp})$ term to the 
denominator of integral kernel. 
In the following we examine two typical cases, namely longitudinal and transverse in-plane bias fields, to see
the behaviors of $\langle\dot{q}\rangle \sim H_{\perp}$ dependence curve.

For longitudinal in-plane bias fields, $H_{\perp}=H_x$ and $\phi_{\perp}=0$. 
For large enough $H_z$, similar calculation provides the new period as
\begin{equation}\label{T_Hz_H_x}
T_x=T_0+\frac{2\pi}{\bar{\gamma} H_z}\frac{\alpha^2\pi^2}{8}\left(\frac{H_x}{H_z}\right)^2,
\end{equation}
thus leads to a new velocity
\begin{equation}\label{v_average_Hz_Hx}
\langle\dot{q}\rangle [H_x]=\langle\dot{q}\rangle_0+\eta\alpha\Delta_0\bar{\gamma} H_z\frac{K_0}{2\pi}\frac{\pi^2}{8}\left(\frac{H_x}{H_z}\right)^2.
\end{equation}
Obviously, for a fixed $H_z$ this $\langle\dot{q}\rangle\sim H_x$ curve
is a parabola going upwards (downwards) with its center locating at $H_x=0$ for $\eta=+1$ ($-1$).

Alternatively for transverse in-plane bias fields, $H_{\perp}=H_y$ and $\phi_{\perp}=\pi/2$. 
Similar calculation provides the high-$H_z$ period as
\begin{equation}\label{T_Hz_H_y}
T_y=T_0+\frac{2\pi}{\bar{\gamma} H_z}\frac{\alpha^2\pi^2}{8}\left[\left(\frac{H_y}{H_z}\right)^2-\frac{\eta K_0}{8\pi}\left(1+\frac{\kappa}{4}\right)\frac{H_y}{H_z}\frac{H_{\mathrm{b}}^0}{H_z}\right].
\end{equation}
The resulting averaged wall velocity is
\begin{equation}\label{v_average_Hz_Hy}
\begin{split}
\langle\dot{q}\rangle [H_y]=& \langle\dot{q}\rangle_0+\eta\alpha\Delta_0\bar{\gamma} H_z\frac{K_0}{2\pi}\frac{\pi^2}{8}\left[\left(\frac{H_y-\delta H_y}{H_z}\right)^2   \right.  \\
& \left. -\left(1+\frac{\kappa}{4}\right)^2\left(\frac{K_0}{2\pi}\right)^2\left(\frac{H_{\mathrm{b}}^0}{H_z}\right)^2 \right],
\end{split}
\end{equation}
with $\delta H_y=\eta\left(1+\frac{\kappa}{4}\right)\frac{K_0}{2\pi}H_{\mathrm{b}}^0$.
Therefore for a fixed $H_z$ and $\eta=+1$ ($-1$), the $\langle\dot{q}\rangle\sim H_y$ curve
becomes a parabola going upwards (downwards) with its center locating at $H_y=\delta H_y$.
For a given magnetic heterostructure with chiral-magnet central layer, the ``hard-easy ratio" $\kappa$
can be calculated. For a chiral DW with certain wall charge $\eta$, by measuring the 
$\langle\dot{q}\rangle\sim H_y$ dependence one can extract out the bDMI strength $D_{\mathrm{b}}$
from the location of parabola center.

Note that in our recent work in Ref. \cite{jlu_PRB_2020}, we have constructed 
a general scheme of identifying and quantifying
bDMI in magnetic heterostructures via precessional flow of chiral DWs under in-plane transverse bias fields.
In that scheme, the linearization of trigonometric functions does not lose too many details of the entire circle
since DWs precess almost evenly under large enough out-of-plane driving fields.
However the ``hard-easy ratio" $\kappa$ is totally neglected since generally people use
the static width $\Delta_0$ instead of the real complicated breathing one. 
This simplification holds for not too narrow magnetic central layers with strong PMA.
However for those with relatively weak PMA and shrinking width, the importance of $\kappa$
will increase significantly.
This effect manifests itself as the additional factor $\left(1+\frac{\kappa}{4}\right)\frac{K_0}{2\pi}$
in our new $\delta H_y$.
We will revisit this issue in the discussion section later.

\section{V. Current-driven dynamics} 
In this section, we turn to in-plane current-driven dynamics of chiral DWs ($H_z=0$ and $j_a\ne 0$).
Parallel deductions will be performed compared with field-driven case in the above section.
Effects of bDMI to both Walker current density and wall behaviors under high currents will be explored.

\subsection{V.A Enlarged Walker current density}
Pioneer works provide us that in the absence of bDMI, the in-plane Walker current density
is $j_{\mathrm{W}}^0=\frac{\Delta_0\gamma H_{\mathrm{W}}^0}{|\alpha-\beta|}\frac{e M_s}{\mu_{\mathrm{B}} P}$.
Without $\mathbf{H}_{\perp}$, the existence condition of rigid-flow mode ($\dot{\varphi}=0$ and $\dot{\Delta}=0$)
turns the first equation in Eq. (\ref{Full_dynamical_equation_bDMI}) to
\begin{equation}\label{g_phi_definition}
\eta\mathrm{sgn}(\alpha-\beta)\frac{j_a}{j_{\mathrm{W}}^0}=g(\varphi)\equiv b\cos\varphi+\frac{\sin 2\varphi}{\sqrt{1+\kappa\sin^2\varphi}}.
\end{equation}
Similarly, for fixed $b$ and $\kappa$ when the maximum absolute value of the function $g$, i.e. $|g(\varphi)|_{\mathrm{max}}$,
is found for $\varphi\in\left[0,2\pi \right)$, the new Walker current density is then obtained as
$j_{\mathrm{W}}=j_{\mathrm{W}}^0 |g(\varphi)|_{\mathrm{max}}$.
Also, the sign of $b$ is irrelevant to $|g(\varphi)|_{\mathrm{max}}$ 
since $g(b,\varphi)=g(-b,\varphi+\pi)$ always holds.
Therefore we can set $b>0$ to make the analysis simple.
For neglectable $\kappa$, $g(\varphi)\equiv f(\varphi,\kappa=0)$ 
thus achieves the same maximum as $|f(\varphi,\kappa=0)|_{\mathrm{max}}$ at the same location.
For finite $\kappa$, after defining $x\equiv \sin\varphi$ and introducing
a new function $\mathcal{G}(x)\equiv [g(\varphi)]^2$, one thus has
$|g(\varphi)|_{\mathrm{max}}=\sqrt{|\mathcal{G}(\varphi)|_{\mathrm{max}}}$.
The extremum condition, $d\mathcal{G}/dx=0$, can be transformed into a quartic equation of $x^2$,
whose exact solution is too complicated to write out explicitly.
Alternatively, by numerically searching the maximum of $\mathcal{G}(x)$ for $x\in [-1,1]$,
the new Walker current density $j_{\mathrm{W}}$ (in the unit of $j_{\mathrm{W}}^0$) and the 
corresponding $\varphi_0$ where $j_{\mathrm{W}}$ is reached are depicted in Fig. 4(a) 
and 4(b), respectively.
For fixed $\kappa$, along with the increase of bDMI $j_{\mathrm{W}}$ is considerably enlarged and
it location $\varphi_0$ decreases from $\pi/4$ to 0 since the first term in Eq. (\ref{g_phi_definition})
becomes dominant.
On the other hand, $\kappa$ hardly changes $j_{\mathrm{W}}$ but strongly affects $\varphi_0$, 
even when bDMI is small. 
For fixed bDMI when $\kappa\gg 1$ the location $\varphi_0$ rapidly decreases from $\pi/4$ to 0.
This comes from the fact that now Eq. (\ref{g_phi_definition}) provides
$g(\varphi)\sim (b+2/\sqrt{\kappa})\cos\varphi$ which achieve maximum absolute value at
$\varphi_0=0$.
In summary, the huge difference between Fig. 3(b) and 4(b) comes from the subtle distinction
between Eqs. (\ref{f_phi_definition}) and (\ref{g_phi_definition}), especially for finite $\kappa$.

\begin{figure} [htbp]
	\centering
	\includegraphics[width=0.43\textwidth]{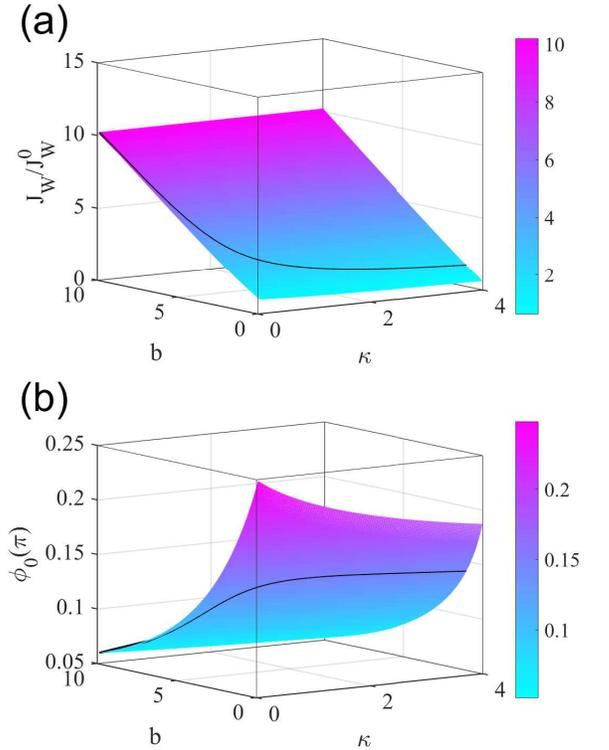}
	\caption{(Color online) (a) New Walker current density $J_{\mathrm{W}}$ in the unit of 
		$j_{\mathrm{W}}^0=\frac{\Delta_0\gamma H_{\mathrm{W}}^0}{|\alpha-\beta|}\frac{e M_s}{\mu_{\mathrm{B}} P}$ and (b) the corresponding in-plane angle under in-plane driving current $j_a\mathbf{e}_x$.
		All data come from direct numerical searching of the maximum of function $\mathcal{G}(x=\sin\varphi)$
		in $(b,\kappa)-$space. The solid curve in (b) comes from $\varphi_0=\arcsin\tilde{x}$ in 
		Eq. (\ref{b2k_4_cubic_solution}), while that in (a) is $\sqrt{\mathcal{G}(\tilde{x})}$.
	}\label{fig4}
\end{figure}

An interesting special case is ``$b^2\kappa=4$", under which
the quartic equation of $x^2$ (coming from $d\mathcal{G}/dx=0$)
is reduced to a cubic equation: ``$\kappa (x^2)^3+(1-2\kappa)(x^2)^2-(4+1/\kappa) x^2 +1=0$".
It's solution
\begin{equation}\label{b2k_4_cubic_solution}
\tilde{x}=\left[\frac{2\kappa-1}{3\kappa}+\frac{4(1+\kappa)}{3\kappa}\cos\frac{\theta+\pi}{3}\right]^{\frac{1}{2}}, \; \theta=\cos^{-1}\frac{\frac{11}{16}-\kappa}{1+\kappa},
\end{equation}
corresponds to the maximum of $\mathcal{G}(x)$. 
The resulting new Walker current density is 
$j_{\mathrm{W}}=j_{\mathrm{W}}^0\sqrt{\mathcal{G}(\tilde{x})}$, which has been plotted in Fig. 4(a)
by a solid curve.
Correspondingly, the in-plane angle $\varphi_0=\arcsin\tilde{x}$ is depicted in Fig. 4(b).
The high coincidence between numerics and analytics of this special case provide strong cross-validation
for both processing methods.
At last for $|j_a|\le j_{\mathrm{W}}$, the wall takes steady-flow mode 
with velocity $\dot{q}=-\frac{\beta}{\alpha} B_J$. 
This means that bDMI does not change the wall's mobility under in-plane currents for steady flows.

\subsection{V.B $\langle\dot{q}\rangle$ under $|j_a|\gg j_{\mathrm{W}}$}
As $|j_a|$ exceeds $j_{\mathrm{W}}$, steady-flow mode fails and the wall undergoes precessional flow.
Under similar approximation and integration over a full circle as in Sec. IV.B, 
the precession period $T'_0$ is
\begin{equation}\label{T_ja_integration_form}
T'_0=\frac{1}{\bar{\gamma} H_{\mathrm{W}}^0 }\int_{0}^{2\pi}\frac{d\varphi/\sqrt{1+\kappa\sin^2\varphi}}{\sigma\frac{j_a}{j_{\mathrm{W}}^0}-\eta\frac{\pi H_{\mathrm{b}}^0}{H_{\mathrm{K}}}\cos\varphi-\frac{\sin 2\varphi}{\sqrt{1+\kappa\sin^2\varphi}}},
\end{equation}
in which $\sigma\equiv \eta\cdot\mathrm{sgn}(\alpha-\beta)$ and a prime in this section
means quantities in current-driven case.
For high-current limit, after preserving the second-order small quantities we have
\begin{equation}\label{T_ja_with_2nd_order_terms}
T'_0\approx\frac{\sigma}{\bar{\gamma} H_{\mathrm{W}}^0 }\frac{j_{\mathrm{W}}^0}{j_a}\left\{K_0 + \left[\frac{K_0}{2}\left(\frac{\pi H_{\mathrm{b}}^0}{H_{\mathrm{K}}}\right)^2+I_1\right]\left(\frac{j_{\mathrm{W}}^0}{j_a}\right)^2\right\},
\end{equation}
in which the integral $I_1(\kappa)$ is defined in Eq. (\ref{I_1_definition}).
Thus the time-averaged wall velocity is
\begin{equation}\label{v_average_ja_with_2nd_order_terms}
\langle\dot{q}\rangle'_0=-\frac{1+\alpha\beta}{1+\alpha^2}B_J+\frac{(\alpha-\beta)B_J}{\alpha(1+\alpha^2)}
\left[\frac{1}{2}\left(\frac{\pi H_{\mathrm{b}}^0}{H_{\mathrm{K}}}\right)^2+\frac{I_1}{K_0}\right]\left(\frac{j_{\mathrm{W}}^0}{j_a}\right)^2.
\end{equation}
If we focus on the velocity dependence on $j_a$ (thus $B_J$), interestingly even in the 
absence of bDMI, our result provides an additional $(j_a)^{-1}$ term except for the well-known
first term on the right hand of Eq. (\ref{v_average_ja_with_2nd_order_terms}).
The appearance of bDMI strengthens this effect.
On the other hand, this term helps to reorganize the $\langle\dot{q}\rangle'_0\sim j_a$ relationship
as $c'(j_a-j_a^0)/j_a+d'/j_a$, which is similar to its counterpart in field-driven 
case [see Eq. (\ref{v_average_Hz_with_2nd_order_terms}) and related discussions].

\subsection{V.C $\langle\dot{q}\rangle \sim H_{\perp}$ dependence under large $j_a$}
The in-plane bias field $\mathbf{H}_{\perp}$ is once again turned on to manipulate chiral DWs' dynamics.
Just similar to what we have discussed in Sec. IV.C, the Waker current density will inevitably be affected by $\mathbf{H}_{\perp}$,
however the exact dependence is hard to obtain.
Now we concentrate on the case in which the in-plane current $j_a$ is sufficiently large
that the wall already falls into the precessional-flow mode with 
the breathing width $\Delta=\Delta_0(1+\kappa\sin^2\varphi)^{-1/2}$.
The period for a full circle is similar to Eq. (\ref{T_ja_integration_form})
except for an additional ``$-\frac{\pi H_{\perp}}{H_{\mathrm{K}}}\frac{\sin(\varphi-\phi_{\perp})}{\sqrt{1+\kappa\sin^2\varphi}}$" term
to the denominator of the integral kernel.

In the first example, we focus on longitudinal in-plane bias fields ($H_{\perp}=H_x$ and $\phi_{\perp}=0$).
Similar calculation shows that the new period is larger than $T'_0$ by an additional term proportional to $(j_a)^{-3}$, that is,
\begin{equation}\label{T_ja_H_x}
T'_x=T'_0+\frac{\sigma I_2}{\bar{\gamma} H_{\mathrm{W}}^0} \left(\frac{\pi H_x}{H_{\mathrm{K}}}\right)^2  \left(\frac{j_{\mathrm{W}}^0}{j_a}\right)^3,
\end{equation}
in which the integral $I_2(\kappa)$ is defined in Eq. (\ref{I_2_definition}). This leads to a new velocity
\begin{equation}\label{v_average_ja_Hx}
\langle\dot{q}\rangle' [H_x]=\langle\dot{q}\rangle'_0+\frac{\alpha-\beta}{\alpha(1+\alpha^2)}
\left(\frac{\pi H_x}{H_{\mathrm{K}}}\right)^2 \frac{I_2}{K_0}\left(\frac{j_{\mathrm{W}}^0}{j_a}\right)^2 B_J.
\end{equation}
For fixed $H_x$, the extra term in the above equation provides extra contribution to the $j_a^{-1}-$term
in wall's velocity [see Eq. (\ref{v_average_ja_with_2nd_order_terms})].
However for fixed $j_a$, the $\langle\dot{q}\rangle' \sim H_x$ curve is a parabola with its 
center locating at $H_x=0$ and its opening direction depends on the relative strength of $\alpha$ and $\beta$. 

Next we turn to transverse in-plane bias fields ($H_{\perp}=H_y$, $\phi_{\perp}=\pi/2$).
Similar calculation yields the large-$j_a$ period as
\begin{equation}\label{T_ja_H_y}
T'_y=T'_0+\frac{\sigma}{\bar{\gamma} H_{\mathrm{W}}^0} \left[\left(\frac{\pi H_y}{H_{\mathrm{K}}}\right)^2 I_3 -2\frac{\pi H_y}{H_{\mathrm{K}}}\frac{\eta\pi H_{\mathrm{b}}^0}{H_{\mathrm{K}}} I_4 \right] \left(\frac{j_{\mathrm{W}}^0}{j_a}\right)^3,
\end{equation}
where the integrals $I_{3,4}(\kappa)$ have also been defined in Eq. (\ref{I_2_definition}).
The resulting averaged wall velocity is then
\begin{equation}\label{v_average_ja_Hy}
\begin{split}
\langle\dot{q}\rangle' [H_y]=& \langle\dot{q}\rangle'_0+\frac{\alpha-\beta}{\alpha(1+\alpha^2)}\left\{\frac{\pi^2 I_3}{K_0}\left(\frac{H_y-\delta H'_y}{H_{\mathrm{K}}}\right)^2   \right.  \\
& \quad \left. -\left(\frac{\pi H_{\mathrm{b}}^0}{H_{\mathrm{K}}}\right)^2\frac{(I_4)^2}{K_0 I_3} \right\}\left(\frac{j_{\mathrm{W}}^0}{j_a}\right)^2 B_J,
\end{split}
\end{equation}
with $\delta H'_y=\eta(I_4/I_3) H_{\mathrm{b}}^0$.
Now the $\langle\dot{q}\rangle \sim H_y$ curve
becomes a parabola with its center locating at $H_y=\delta H'_y$.
Also, its opening direction has nothing to do with the wall's topological charge,
but only depends on $\mathrm{sgn}(\alpha-\beta)$.
For a given magnetic heterostructure with chiral-magnet central layer, 
the ``hard-easy ratio" $\kappa$ (thus $I_3$, $I_4$) can be calculated. 
Similar to field-driven case, for a chiral DW with a certain wall charge $\eta$, by measuring the
$\langle\dot{q}\rangle' \sim H_y$ dependence one can also extract out 
the bDMI strength $D_{\mathrm{b}}$ from the location of parabola center.
The data from field- and current-driven cases can be cross-checked to confirm the value of bDMI
strength in the underlying magnetic heterostructure.

\section{VI. Discussions} 
First of all, we want to address the feasibility of the classical Walker ansatz 
in Eq. (\ref{q_phi_Delta_ansatz}) adopted in this work.
In perfect strip-shaped heterostructures, early studies show that DMIs can induce the wall tiling
$\chi$ with respect to $+\mathbf{e}_y$\cite{Boulle_PRL_2013}.
However for real heterostructures with disorders, the walls take complex meandering shape with its magnetization
vector rotating several times along the wall thus show inconspicuous tilting \cite{Pizzini_EPL_2016,Pizzini_PRB_2016,Marrows_PRB_2018}. 
This leads to negligible longitudinal component of $\mathbf{H}_{\mathrm{b}}$ 
which is proportional to $\nabla_y m_z$,
hence explains the feasibility of using 1D Walker ansatz.
Another neglected effect is the magnetization canting $\theta_{\infty}$\cite{jlu_PRB_2016} in domains 
by in-plane fields either from intrinsic bDMI or from external exertion.
When both $\chi$ and $\theta_{\infty}$ are considered, a more complicated wall ansatz
\begin{equation}\label{q_phi_chi_Delta_ansatz}
\tan\frac{\vartheta}{2}=\frac{e^R+\tan(\theta_{\infty}/2)}{1+e^R \tan(\theta_{\infty}/2)},\quad \phi=\varphi(t)
\end{equation}
can be proposed with $R\equiv \eta[(x-q)\cos\chi+y\sin\chi]/\Delta$.
By integrating the resulting dynamical equations over infinite strip length and finite width, 
the so-called ``$q-\varphi-\chi$"\cite{Boulle_PRL_2013} or ``$q-\varphi-\chi-\Delta$"\cite{Nasseri_JMMM_2017,Nasseri_JMMM_2018} models, emerge.
However they are too complicated to provide clear physical pictures in analyzing chiral wall dynamics. 

Second, in steady-flow mode the wall width has explicit expression.
While in precessional-flow mode the wall begins to breath, leading to a time-dependent wall width.
The simple approximation in Sec. IV and V, i.e. $\Delta=\Delta_0(1+\kappa\sin^2\varphi)^{-1/2}$,
is directly fetched from the steady-flow mode and is not the exact solution of $\Delta(t)$.
However in most cases the wall does not change too much in a full circle ($|\dot{\Delta}/\Delta|\ll 1$).
Therefore it can be regarded as a good approximate description of the actual wall width.

Third, under sufficiently large out-of-plane fields or in-plane currents, the chiral walls fall into
the precessional-flow mode. In this work, during a full circle ($0\le \varphi <2\pi$) 
the ``Taylor expansion" method
is adopted to get higher order correction (here we preserve to the second order).
Also, the role of ``hard-easy ratio" $\kappa$ is fully revealed,
especially in the center offsets of $\langle\dot{q}\rangle [H_y]$ and $\langle\dot{q}\rangle' [H_y]$ parabolas.
This strategy holds under the assumption that magnetic anisotropic, in-plane bias and bDMI effective fields 
are all small compared with out-of-plane driving fields or in-plane currents (through $j_{\mathrm{W}}^0$
which is proportional to $H_{\mathrm{W}}^0$).
Generally, this condition is not hard to achieve thus makes the corresponding measurements feasible.
However for chiral magnets with sufficient large bDMI, to achieve the full parabola quite large
driving fields or currents have to be exerted which may make the structure of chiral domain walls unstable.
This possibility limits the application of our theories presented above.

Alternatively, in our recent work (see Ref. \cite{jlu_PRB_2020}) another approximation 
has been adopted: for large enough $H_z$ or $j_a$,
DWs precess almost evenly in a full circle.
After linearization of trigonometric functions in dynamical equations,
the average wall velocity within $\varphi\in[0,1)$ is used to mimic the one over a full circle.
Regarding this approach, we would like to present several comments:
(i) The wall width is always taken as the static one, $\Delta_0$, which is $\kappa$-independent.
This may not affect too much for wide magnetic heterostructures, however in relatively narrow
ones the effects of $\kappa$ could get stronger.
(ii) This approach is not subject to the limitation that all other fields should be small compared with
out-of-plane driving fields or in-plane currents, however it suffers from the 
constraint that analytics can only hold not too far away from the dome summits or canyon bottoms.
Therefore, it can not explain the further evolution of wall velocity when in-plane bias fields 
go faraway from the centers of domes or canyons.
(iii) After series expansions, an additional absolute linear term emerges 
which is the direct consequence of linearization operation.
For example, to compare with Eqs. (\ref{v_average_Hz_with_2nd_order_terms}) and (\ref{v_average_Hz_Hy}) 
in the present work, the field-driven wall velocity $v_{\mathrm{b,T}}$ in Ref. \cite{jlu_PRB_2020} 
can be expanded as
\begin{equation}\label{v_b_T_expansion}
\frac{v_{\mathrm{b,T}}}{\eta\alpha\Delta_0\bar{\gamma} H_z}=1+\frac{\pi}{4\alpha}|\Gamma|+\frac{\pi^2}{12}|\Gamma|^2,
\end{equation}
with $\Gamma\equiv \frac{H_y}{H_z}-\eta\frac{H_{\mathrm{b}}^0}{H_z}-\frac{2}{\pi}\frac{H_{\mathrm{K}}}{H_z}$.
Therefore the domes or canyons in Ref. \cite{jlu_PRB_2020} are generally not parabolas but cones
around the dome summits or canyon bottoms.
However, considering the fact that $\eta\left(1+\frac{\kappa}{4}\right)$ and $I_4/I_3$ approach 1 
when $\kappa\rightarrow 0$ and $H_{\mathrm{K}}/H_z$ becomes neglectable for not-too-narrow geometries,
the correctness of both two analytical schemes can be cross-verified.

\section{acknowledgments} 
M.L. acknowledges support from the National Natural Science Foundation of China (Grant No. 11947023).
B.X. is supported by the National Natural Science Foundation of China (Grant No. 11774300).


\appendix

\section{Appendix A: Maximum value and location of $|f(\varphi)|$}
\setcounter{equation}{0}
\renewcommand{\theequation}{A\arabic{equation}}

By setting $x\equiv \sin\varphi$ and defining
\begin{equation}\label{Capital_F_definition}
\mathcal{F}(x)\equiv [f(\varphi)]^2 = \left(1-x^2\right)\left(2x+b\sqrt{1+\kappa x^2}\right)^2,
\end{equation}
the searching of maximum of $|f(\varphi)|$ for $\varphi\in\left[0,2\pi \right)$ is equivalent to 
the counterpart of $\mathcal{F}(x)$ for $|x|\le 1$.
We denote the location of maximum point as $\tilde{x}$ and the corresponding maximum of $\mathcal{F}(x)$
is $\mathcal{F}(x)|_{\mathrm{max}} = \left(1-\tilde{x}^2\right)\left(2\tilde{x}+b\sqrt{1+\kappa \tilde{x}^2}\right)^2$.
The resulting maximum of $|f(\varphi)|$ is thus $\sqrt{\mathcal{F}(x)|_{\mathrm{max}}}$ at
$\sin\varphi_0=\tilde{x}$.

Standard calculus provides us the following results:

\noindent
[a] when $0<\kappa \le 1$:

\noindent
(a1) for $0<b^2<4/\kappa$:
	\begin{equation}\label{x_tilde_a1}
	\begin{split}
	\tilde{x} &=\left[\frac{\kappa-1}{3\kappa}+\sqrt{\frac{4\left(\kappa+2\right)^2-b^2\kappa(1-\kappa)^2}{9\kappa^2(4-b^2\kappa)}}\cdot\cos\frac{\theta+\pi}{3}\right]^{\frac{1}{2}},  \\
\theta &=\arccos\left\{\frac{\sqrt{1-\frac{b^2\kappa}{4}}\cdot\left[1+\frac{b^2\kappa}{4}\left(\frac{1-\kappa}{\kappa+2}\right)^3\right]}{\left[1-\frac{b^2\kappa}{4}\left(\frac{1-\kappa}{\kappa+2}\right)^2\right]^{3/2}}\right\},
	\end{split}
	\end{equation}

\noindent
(a2) for $b^2=4/\kappa$:
\begin{equation}\label{x_tilde_a2}
\tilde{x}=\sqrt{\frac{\kappa}{2\kappa+1}},
\end{equation}

\noindent
(a3) for $b^2>4/\kappa$:
\begin{equation}\label{x_tilde_a3}
\begin{split}
\tilde{x} &=\sqrt{-\frac{1-\kappa}{3\kappa}-\frac{\sqrt[3]{Y_+}+\sqrt[3]{Y_-}}{3\kappa(4-b^2\kappa)}},  \\
Y_{\pm} &=(b^2\kappa-4)^2\left\{\frac{(\kappa+2)^3}{2} +\frac{b^2\kappa}{8}(1-\kappa)^3  \right.   \\
& \quad \left. \pm\frac{3\kappa}{2}\sqrt{\frac{3b^2\left[(\kappa+2)^3+\frac{b^2}{4}(\kappa+1)(1-\kappa)^3\right]}{b^2\kappa-4}}\right\},
\end{split}
\end{equation}

\noindent
[b] when $\kappa>1$:

\noindent
(b1) for $0<b^2<4/\kappa$: same as Eq. (\ref{x_tilde_a1}),

\noindent
(b2) for $b^2=4/\kappa$: same as Eq. (\ref{x_tilde_a2}),

\noindent
(b3) for $4/\kappa<b^2<4(\kappa+2)^3/[(\kappa+1)(\kappa-1)^3]$: same as Eq. (\ref{x_tilde_a3}),

\noindent
(b4) for $b^2=4(\kappa+2)^3/[(\kappa+1)(\kappa-1)^3]$:
\begin{equation}\label{x_tilde_b4}
\tilde{x}=\sqrt{\frac{\kappa^2-1}{\kappa(2\kappa+1)}},
\end{equation}

\noindent
(b5) for $b^2>4(\kappa+2)^3/[(\kappa+1)(\kappa-1)^3]$:
\begin{equation}\label{x_tilde_b5}
\begin{split}
\tilde{x} &=\left[\frac{\kappa-1}{3\kappa}+\sqrt{\frac{b^2\kappa(\kappa-1)^2-4\left(\kappa+2\right)^2}{9\kappa^2(b^2\kappa-4)}}\cdot\cos\frac{\theta}{3}\right]^{\frac{1}{2}},  \\
\theta &=\arccos\left\{-\frac{\sqrt{1-\frac{4}{b^2\kappa}}\cdot\left[1-\frac{4}{b^2\kappa}\left(\frac{\kappa+2}{\kappa-1}\right)^3\right]}{\left[1-\frac{4}{b^2\kappa}\left(\frac{\kappa+2}{\kappa-1}\right)^2\right]^{3/2}}\right\}.
\end{split}
\end{equation}

\section{Appendix B: Definitions and values of several integrals}
\setcounter{equation}{0}
\renewcommand{\theequation}{B\arabic{equation}}
In this appendix, several integrals appeared in the main text
are listed and calculated.

The first one is $K_0$ appearing in Sec. IV.B and throughout this paper:
\begin{equation}\label{K_0_definition}
K_0=\int_{0}^{2\pi}\frac{d\varphi}{\sqrt{1+\kappa\sin^2\varphi}}=\frac{4}{\sqrt{1+\kappa}}K\left(\sqrt{\frac{\kappa}{1+\kappa}}\right),
\end{equation} 
in which $K(k)\equiv\int_{0}^{\pi/2} \frac{d\omega}{\sqrt{1-k^2\sin^2\omega}}$ is the complete elliptic integral of the first kind.

The second one is $I_1$ which appeared in Sec. V.B:
\begin{equation}\label{I_1_definition}
\begin{split}
I_1=&\int_{0}^{2\pi}\frac{\sin^2 2\varphi}{(1+\kappa\sin^2\varphi)^{3/2}}d\varphi   \\
   =&\frac{16\sqrt{1+\kappa}}{\kappa}\left[\frac{2+\kappa}{\kappa(1+\kappa)}K\left(\sqrt{\frac{\kappa}{1+\kappa}}\right)-\frac{2}{\kappa}E\left(\sqrt{\frac{\kappa}{1+\kappa}}\right)\right],
\end{split}
\end{equation} 
in which $E(k)\equiv\int_{0}^{\pi/2}d\omega \sqrt{1-k^2\sin^2\omega} $ is the complete elliptic integral of the second kind.

The next three integrals are $I_2$, $I_3$ and $I_4$ which appeared in Sec. V.C:
\begin{equation}\label{I_2_definition}
\begin{split}
I_2=&\int_{0}^{2\pi}\frac{\sin^2 \varphi}{(1+\kappa\sin^2\varphi)^{3/2}}d\varphi   \\
=&\frac{4}{\kappa\sqrt{1+\kappa}}\left[K\left(\sqrt{\frac{\kappa}{1+\kappa}}\right)-E\left(\sqrt{\frac{\kappa}{1+\kappa}}\right)\right], \\
I_3=&\int_{0}^{2\pi}\frac{\cos^2 \varphi}{(1+\kappa\sin^2\varphi)^{3/2}}d\varphi   \\
=&\frac{4}{\kappa\sqrt{1+\kappa}}\left[(1+\kappa)E\left(\sqrt{\frac{\kappa}{1+\kappa}}\right)-K\left(\sqrt{\frac{\kappa}{1+\kappa}}\right)\right], \\
I_4=&\int_{0}^{2\pi}\frac{\cos^2 \varphi}{1+\kappa\sin^2\varphi}d\varphi   =2\pi\frac{\sqrt{1+\kappa}-1}{\kappa}.
\end{split}
\end{equation}


\end{document}